# Two types of magnetic bubbles in MnNiGa observed via Lorentz microscopy

Hiroshi Nakajima[1], Atsuhiro Kotani[1], Ken Harada[1,2], and Shigeo Mori[1, *]

[1]Department of Materials Science, Osaka Prefecture University, Sakai, Osaka 599-8531, Japan

[2]Center for Emergent Matter Science (CEMS), The Institute of Physical and Chemical Research (RIKEN), Hatoyama, Saitama 350-0395, Japan

*E-mail: mori@mtr.osakafu-u.ac.jp

ABSTRACT

Magnetic bubbles are remarkable spin structures that developed in uniaxial magnets with strong magnetocrystalline anisotropy. Several contradictory reports have been published concerning the magnetic bubble structure in a metallic magnet MnNiGa: Biskyrmions or type-II bubbles. Lorentz microscopy in polycrystalline MnNiGa was used to explain the magnetic bubble structure. Depending on the connection between the magnetic easy axis and the observation plane, two types of magnetic bubbles were formed. Magnetic bubbles with 180° domains were formed if the easy axis was away from the direction perpendicular to the observation plane. The contrast of biskyrmion is reproduced by this form of a magnetic bubble. When the easy axis was approximately perpendicular to the observing plane, type-II bubbles were observed in the same specimen. The findings will fill a knowledge gap between prior reports on magnetic bubbles in MnNiGa.

# 1. Introduction

Magnetic bubbles are circularly rotating magnetic structures formed in uniaxial magnets with high magnetocrystalline anisotropy. The structures have been known for more than 50 years in ferromagnets such as cobalt and hexagonal ferrites.[1,2] Nevertheless, the recent finding of magnetic skyrmions has rekindled interest in magnetic bubbles, as magnetic bubbles have a topological number comparable with magnetic skyrmions.[3,4] When the spin structure spans the entire sphere, the topological number $N$ is specified as $N = 1$. As a result of the interest, various unusual magnetic bubbles have been discovered in various magnets, including perovskite manganese oxides and layered oxides.[5–7] Moreover, it is demonstrated that magnetic bubbles can be moved by a current.[8] The topological number is used to classify magnetic bubbles. A magnetic bubble is classified as type-I if it has a continuous rotating structure with $N = 1$. Type-II bubbles, conversely, have two parallel domain walls ($N = 0$). Bloch lines are formed as two discontinuous points in type-II magnetic bubbles.[9]

MnNiGa is reported to form magnetic bubbles.[10] The material has a layered $Ni_2In$-type structure of hexagonal symmetry (space group $P6_3/mmc$). According to the paper, the magnetic bubbles are biskyrmions, which are made up of clockwise and counterclockwise bubbles with $N = 2$. Recent research, however, has called into question the existence of biskyrmion. A Lorentz microscopy study reports that only type-I and type-II magnetic bubbles are found in MnNiGa and that biskyrmions can be explained by type-II magnetic bubbles.[11] A further x-ray holography analysis of MnNiGa bubble formations determined that the magnetic bubbles are type-II.[12] Additionally, Lorentz imaging research based on the transport of intensity equation (TIE) shows that type-II bubbles resemble biskyrmions when TIE parameters are incorrect.[13] Although magnetic domain structures observed by these authors are demonstrated to be type-II in MnNiGa, the contrast of biskyrmons[10] is different from that predicted from the type-II bubble structure because the contrast has no Bloch lines. Moreover, the domain walls of a uniaxial magnet should contain a pair of bright and dark Fresnel contrast lines in type-II magnetic bubbles. The domain walls of biskyrmions, conversely, have just single bright or dark lines in the Fresnel image and no pair structure is visible in the contrast.[10]

We observed magnetic domains in MnNiGa to understand the difference between the magnetic bubble structures reported in the previous studies[10–13]. When the magnetic easy axis of the $c$ axis was away from the direction perpendicular to the viewing plane, magnetic bubbles with 180° domains were created (The plane is the surface formed during the thinning process). When the TIE method was used on this structure, it created the biskyrmion contrast. Type-II magnetic bubbles were also observed when the $c$ axis was slightly inclined from the perpendicular direction to the observing plane. These two structures were discovered in various grains of the same material. The different crystallographic directions concerning the observation planes should explain the discrepancy of the previous studies.

## 2. Experimental methods

Arc melting was used to construct polycrystal specimens of MnNiGa. The ingots of Ga and the powers of Mn and Ni were weighed on the basis of the composition. They were melted in an argon atmosphere using an arc furnace. They were sealed off in a vacuum with a quartz tube. Then, it was heated at 700°C for 110 h. Finally, it was cooled to room temperature by being immersed in water. The specimens were confirmed by x-ray diffraction and energy-dispersive x-ray spectroscopy (EDS) of scanning electron microscopy. The EDS results showed that the ratio of the specimen was Mn:Ni:Ga = 0.98:1.10:1.00, which agrees with the composition. A vibrating sample magnetometer was used to measure magnetization (Quantum Design, Inc.). Lorentz microscopy was conducted using a JEM-2100F transmission electron microscope (TEM) (JEOL Co. Ltd., Japan). The magnitude of the acceleration voltage was 200 kV. Magnetic domain walls were detected using Fresnel imaging. A defocus value $\Delta f$ of Fresnel imaging is defined as positive in an overfocus condition. Small-angle electron diffraction (SmAED) patterns were observed to analyze a magnetic domain structure.[14),15)] The objective lens of the TEM was used to apply external magnetic fields. The images based on TIE were calculated using software (QPt).[16)]

## 3. Results and discussion

Figure 1(a) shows the powder x-ray diffraction profile of the specimen synthesized in this study. The specimen can be indexed with the MnNiGa structure of the space group $P6_3/mmc$, whose structure is illustrated in Figure 1(b). The temperature dependence of magnetization shows a ferromagnetic transition at 350 K [Figure 1(c)]. These results are consistent with those of the previous study.[10)]

Lorentz microscopy was used to examine magnetic domains in the MnNiGa material. Figure 2(a) illustrates the Fresnel image at room temperature in the absence of a magnetic field. The inset electron diffraction pattern reveals that this location has a zone axis of $[2\bar{2}1]$. The simulation of Supplementary Figure 1 indicates that the $c$ axis almost points parallel to the specimen surface. Hence, the $c$ axis of the magnetic easy axis was tilted from the direction perpendicular to the observation plane by more than 10°. Striped domains were formed in this grain although the remanent magnetic field of the objective lens caused some magnetic bubbles. A branching maze pattern is typical of a uniaxial magnet[17)]. Magnetization points the in-plane directions in the domains and the out-of-plane directions at the domain walls, as shown in the magnified image of Figure 2(b): The formation of 180° domains in this region is shown in red arrows. The contrast of Fresnel imaging explains this: When the domain structure is 180° domains, single bright or dark lines are observed at domain walls, as shown in Figure 2(b). If the magnetization in domains points in out-of-plane directions, a pair of bright and dark lines should form at the walls. The SmAED pattern of Figure 2(c) also supports the 180° domain structure depicted by the arrows. The magnetic deflection spots ($\alpha$ = 17.8 μrad) correspond to a width of a pair of antiparallel domains ($d$ = 145 nm) because these values satisfy Bragg's law $d\sin\alpha = \lambda$, where $\lambda$ = 2.51 pm is the wavelength of electrons. When viewed from the direction parallel to the magnetic easy axis, a uniaxial magnet often displays striped domains with out-of-plane magnetization. Nevertheless, as shown in this area, 180° domains arise when the magnetic easy axis is positioned away from the direction perpendicular to the observation plane.

To confirm the formation process of magnetic bubbles, a magnetic field was applied in this region. Figure 3(a) shows the Fresnel image of magnetic bubbles at 300 mT at the area where the transition from the stripe to bubble structures occurred. When an external magnetic field was generated parallel to the TEM's optical axis, domains parallel to the field increased while domains antiparallel to the field decreased. The antiparallel domains were then reduced and shortened on the basis of the strength of the external magnetic field. Consequently, the domains with a pair of bright and dark regions were formed as magnetic bubbles that the domains were pinched off. The magnetization directions are represented in Figure 3(b). Across the domain barriers, the magnetization directions are antiparallel. The structure suggests that the magnetic bubbles originated by shrinking the 180° domains. A similar color map is generated in the TIE mapping[10], as shown in Figure 3(c). The magnetic bubbles formed by pinching off 180° domains depict the contrast of a pair of dark and bright half circles in the Fresnel image as shown in Figure 3(d). On the basis of the above results, the contrast can be explained by the antiparallel magnetization of the red arrows, as illustrated in Figure 3(e).

Another grain with the *c* axis almost perpendicular to the viewing plane was viewed to compare the previous results. The electron diffraction pattern in Figure 4(a) showed the [001] zone axis by tilting a few degrees in this location. Figure 4(b)–(d) illustrates the formation of magnetic bubbles in magnetic fields. Note that the condition of the external magnetic field direction remained unchanged: The tilt of the specimen holder was the same between Figures 3 and 4 although the different grains had different directions of the *c* axis in terms of the observation plane. Striped domains were generated in the absence of a magnetic field. The magnetic field, conversely, generated magnetic bubbles with walls made up of a pair of bright and dark lines. Moreover, the contrast is reversed at two points known as Bloch lines. Consequently, the images revealed that the magnetic bubbles are type-II. This type-II bubble structure is seen in the magnified image of Figure 4(e), which corresponds to the contrast schematic in Figure 4(f). Within the magnetic domains, the magnetization points along the observation direction. The Bloch walls have in-plane magnetization as shown by the red arrows. This model of Figure 4(f) is different from that of Figure 3(e) in terms of the relationships between the magnetization direction and the observation plane. The type-II magnetic bubbles are a typical characteristic of uniaxial magnets when the external magnetic fields are applied from the direction slightly tilted from the magnetic easy axis[18].

The difference between these two structures can be better understood by looking at through-focus images. Through-focus images of magnetic bubbles generated by 180° domains, which reproduce the contrast of biskyrmions, are shown in Supplementary Figure 2. If the defocus value is reversed in the images, the contrast of the magnetic bubbles is changed. The contrast disappears in the in-focus $\Delta f = 0$ μm. The contrast of the magnetic bubbles is higher if the defocus value is larger. Noticeably, the magnetic domain contrast remained unchanged even if the defocus value is changed: Magnetic bubbles are shown by a pair of bright and dark patches in the Fresnel images of the large ($\Delta f = \pm 960$ μm) and small ($\Delta f = \pm 190$ μm) defocus values. Through-focus images of type-II magnetic bubbles (Supplementary Figure 3), on the other hand, demonstrate a defocus dependence of the Fresnel contrast. In the low defocus values ($\Delta f = \pm 190$ μm and $\pm 380$ μm), the magnetic domain walls comprise a pair of bright and dark lines, as illustrated in Figure 4(f). When the defocus is large, the contrast of domain walls merges. For example, the blur contrast in the Fresnel image of $\Delta f = +1400$ μm is similar to the magnetic domain contrast in Supplementary Figure 2. These findings show that magnetic bubbles with 180° domains and type-II may be differentiated using through-focus imaging.

In MnNiGa, there has been controversy in the existence of biskyrmions.[11]–[13] In the previous studies, the authors claim that biskyrmions are the same as type-II magnetic bubbles. They demonstrated that the magnetic structure observed in a MnNiGa single crystal can be experimentally explained by type-II magnetic bubbles in Fresnel imaging and x-ray holography. However, the contrast of biskyrmions in the prior study's Fresnel image[10] appears to be distinct from that of type-II magnetic bubbles. We were able to see two types of magnetic bubbles in the specimen during our examination. The angles formed by the magnetic easy axis and the viewing plane determined them. Magnetic bubbles are 180° domains that are pinched off by the external field when the easy axis is away from the direction perpendicular to the observing plane, reproducing the contrast of biskyrmion. The formation of 180° domains was reasonable because the easy axis was almost parallel to the observation plane. Conversely, type-II magnetic bubbles were also observed when the *c* axis was slightly tilted from the direction perpendicular to the observation plane. They were observed under the same conditions, such as the direction of the external magnetic fields and the specimen tilt, although the directions of the *c* axis differed due to grain differences. In a previous study[10], the authors also demonstrated that a specimen tilt of 25° caused biskyrmion contrast, most likely because the tilt's in-plane component of the magnetic field changed type-II magnetic bubbles into 180° domains. However, the current results show that if a polycrystalline specimen is observed, the specimen tilt is not required to observe the contrast of biskyrmion. The findings of our observations should help to explain the various assertions made in previous studies. Notably, the contrast of Figure 3 is derived from 180° domains that are pinched off by the magnetic fields. Hence, the contrast of Figure 3 should be formed in a ferromagnetic phase under external magnetic fields, which may explain the wide temperature range of the contrast as reported.[10] The reported 180° domains that are pinched off are frequent properties of uniaxial magnets, as observed in another uniaxial magnet,[19] when the magnetic easy axis is more than 10° away from the observing direction.

## 4. Conclusions

Lorentz microscopy was used in this research to observe magnetic bubbles in MnNiGa. We were able to observe two sorts of magnetic bubbles on the basis of the magnetic easy axis directions. One type is magnetic bubbles originating from shortened 180° domains, which reproduce the contrast of biskyrmion. The other is type-II magnetic bubbles with two Bloch lines, which are common in uniaxial magnets. The observations show that the two types of magnetic bubbles can occur in MnNiGa, which explains the contradiction in previous research.

**References**

1) P. J. Grundy, D. C. Hothersall, G. A. Jones, B. K. Middleton and R. S. Tebble, Phys. status solidi **9**, 79 (1972).
2) P. J. Grundy and S. R. Herd, Phys. status solidi **20**, 295 (1973).
3) S. Mühlbauer, B. Binz, F. Jonietz, C. Pfleiderer, A. Rosch, A. Neubauer, R. Georgii and P. Böni, Science **323**, 915 (2009).
4) N. Nagaosa and Y. Tokura, Nat. Nanotechnol. **8**, 899 (2013).
5) A. Kotani, H. Nakajima, K. Harada, Y. Ishii and S. Mori, Phys. Rev. B **94**, 024407 (2016).
6) D. Morikawa, X. Z. Yu, Y. Kaneko, Y. Tokunaga, T. Nagai, K. Kimoto, T. Arima and Y. Tokura, Appl. Phys. Lett. **107**, 212401 (2015).
7) A. Kotani, H. Nakajima, K. Harada, Y. Ishii and S. Mori, Phys. Rev. B **95**, 144403 (2017).
8) W. Jiang, P. Upadhyaya, W. Zhang, G. Yu, M. B. Jungfleisch, F. Y. Fradin, J. E. Pearson, Y. Tserkovnyak, K. L. Wang, O. Heinonen and others, Science **349**, 283 (2015).
9) K. Kurushima, K. Tanaka, H. Nakajima, M. Mochizuki and S. Mori, J. Appl. Phys. **125**, 053902 (2019).
10) W. Wang, Y. Zhang, G. Xu, L. Peng, B. Ding, Y. Wang, Z. Hou, X. Zhang, X. Li, E. Liu and others, Adv. Mater. **28**, 6887 (2016).
11) J. C. Loudon, A. C. Twitchett-Harrison, D. Cortés-Ortuño, M. T. Birch, L. A. Turnbull, A. Štefančič, F. Y. Ogrin, E. O. Burgos-Parra, N. Bukin, A. Laurenson and others, Adv. Mater. **31**, 1806598 (2019).
12) L. A. Turnbull, M. T. Birch, A. Laurenson, N. Bukin, E. O. Burgos-Parra, H. Popescu, M. N. Wilson, A. Stefančič, G. Balakrishnan, F. Y. Ogrin and others, ACS Nano **15**, 387 (2020).
13) Y. Yao, B. Ding, J. Cui, X. Shen, Y. Wang, W. Wang and R. Yu, Appl. Phys. Lett. **114**, 102404 (2019).
14) H. Nakajima, A. Kotani, K. Harada, Y. Ishii and S. Mori, Microscopy **65**, 473 (2016).
15) H. Nakajima, A. Kotani, K. Harada and S. Mori, Microscopy **67**, 207 (2018).
16) K. Ishizuka and B. Allman, J. Electron Microsc. (Tokyo). **54**, 191 (2005).
17) H. Nakajima, H. Kawase, K. Kurushima, A. Kotani, T. Kimura and S. Mori, Phys. Rev. B **96**, 024431 (2017).
18) H. Nakajima, A. Kotani, K. Harada, Y. Ishii and S. Mori, Phys. Rev. B **94**, 224427 (2016).
19) T. Nagai, M. Nagao, K. Kurashima, T. Asaka, W. Zhang and K. Kimoto, Appl. Phys. Lett. **101**, 162401 (2012).

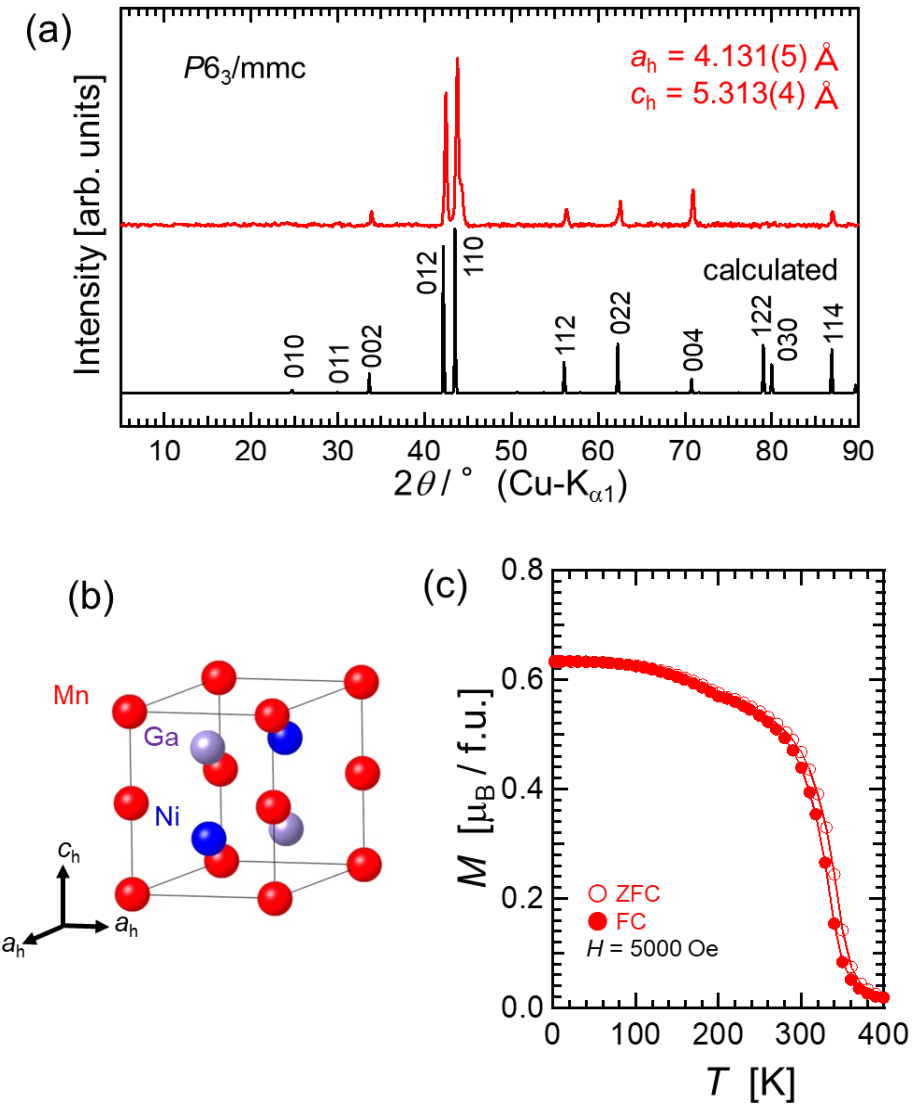

Figure 1. (a) X-ray diffraction profile in MnNiGa. The index is based on the $P6_3/mmc$ structure. The calculated lattice constants were listed. (b) Crystal structure of MnNiGa. (c) The temperature dependence of magnetization. The solid and hollow circles were measured in the field- and zero field-cooling conditions, respectively.

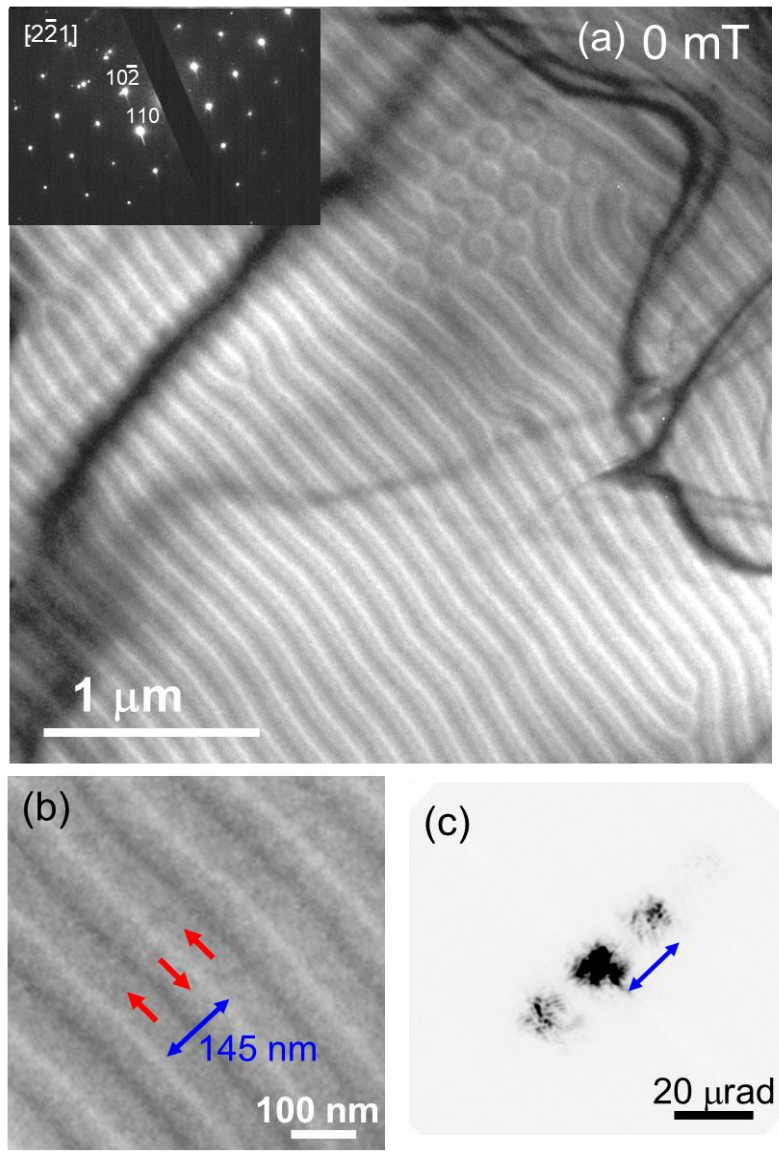

Figure 2. (a) Fresnel image (overfocus $\Delta f = +190$ μm). The inset is the electron diffraction pattern in this area. (b) Magnified image of the magnetic domains. (c) Small-angle electron diffraction pattern from magnetic domains in MnNiGa. The camera length used in the diffraction pattern was approximately 240 m.

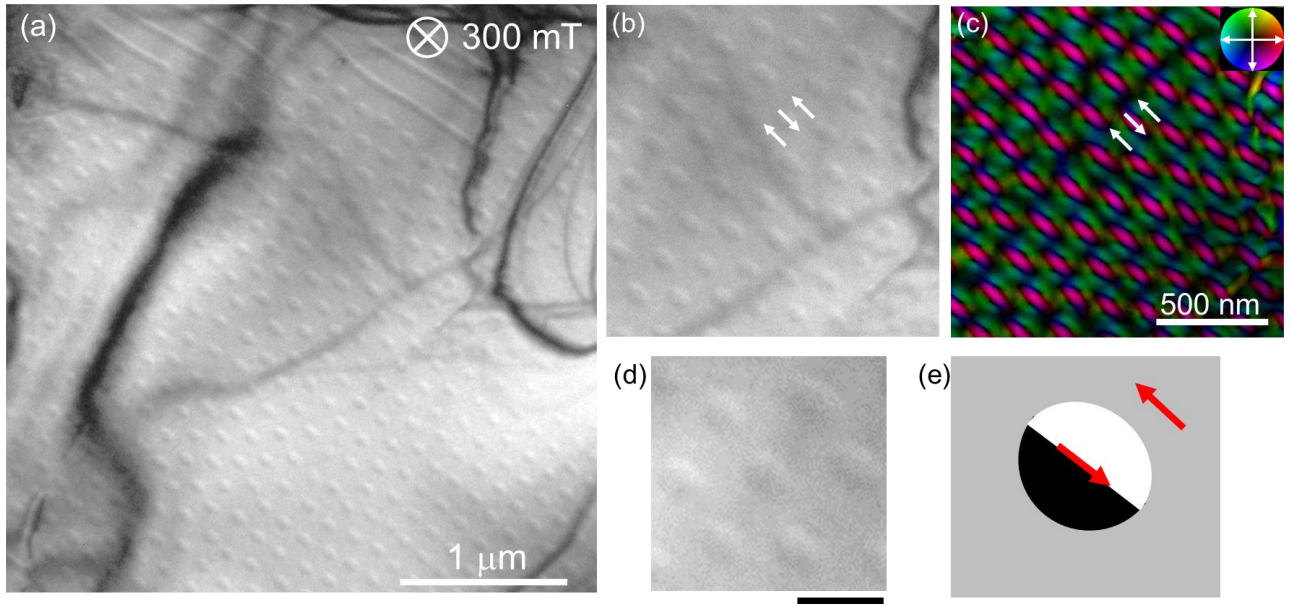

Figure 3. (a) Magnetic domains under the external magnetic field of 300 mT. The magnetic bubbles were formed by pinching off the 180° domains. The defocus value was $\Delta f = -380$ μm (underfocus). (b) Magnified Fresnel image of magnetic domains and (c) magnetization maps based on the transport of intensity equation. The arrows indicate the directions of magnetization. In this area, the *c* axis was away from the observation direction: The angle difference was more than 10°. (d) Magnified image of magnetic bubbles. The scale bar is 200 nm. (e) Schematic of the contrast of a magnetic bubble that is pinched off from 180° domains. The red arrows represent the magnetization direction.

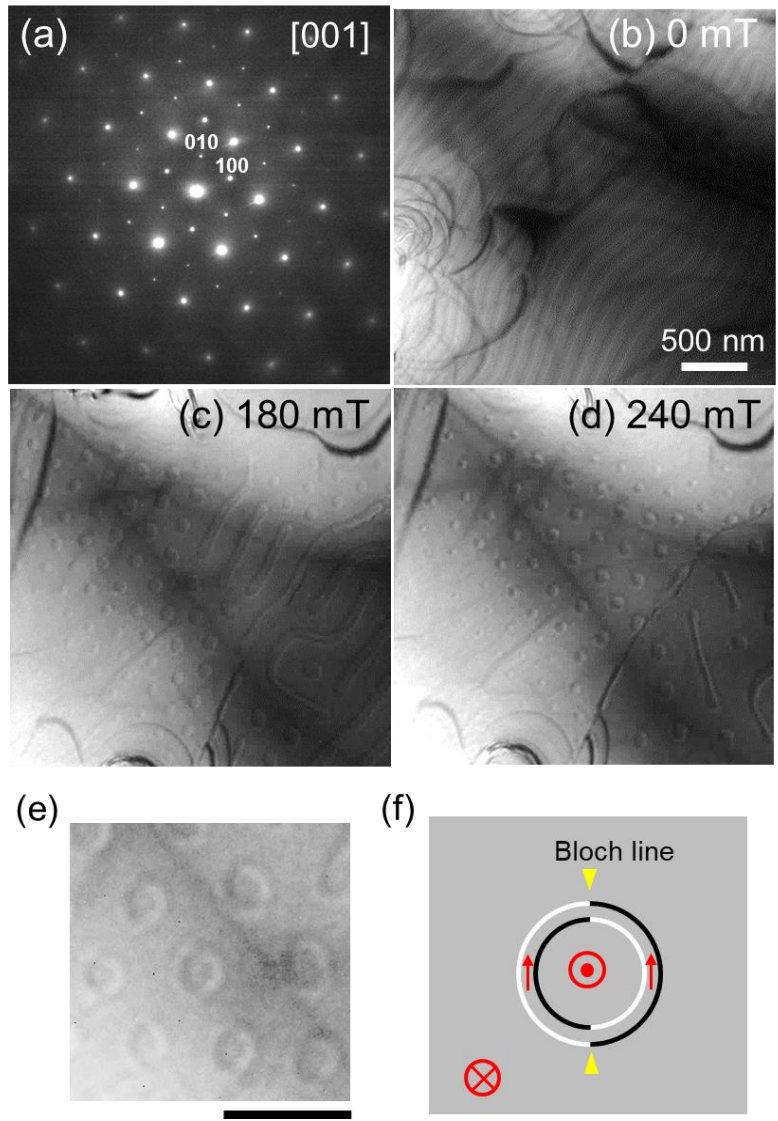

Figure 4. Formation of type-II magnetic bubbles. (a) Electron diffraction pattern. (b) Striped magnetic domains without a magnetic field. (c) The transition from the striped domains to magnetic bubbles at 180 mT. Further increase in the magnetic field at 240 mT increased magnetic bubbles in (d). In this area, the *c* axis was slightly tilted from the observation direction (approximately 3°). The defocus value was $\Delta f = -380$ μm (underfocus). (e) Magnified image of magnetic bubbles. The scale bar is 200 nm. (f) Schematic of the contrast of type-II magnetic bubbles. The red arrows represent the magnetization directions. The yellow arrowheads indicate Bloch lines.

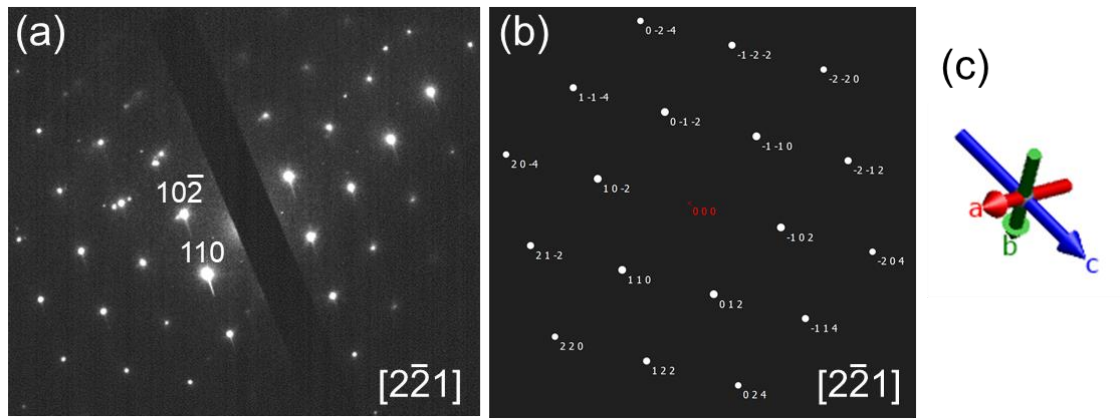

**Supplementary Figure 1. Comparison of the diffraction pattern and simulation.** (a) The electron diffraction obtained from the area of Figure 2. (b) Simulation of an electron diffraction pattern along the $[2\bar{2}1]$ direction. The indexes are shown alongside the diffraction spots. The simulation is based on the $P6_3/mmc$ structure. (c) The directions of the unit cell viewed along the $[2\bar{2}1]$ direction. This simulation demonstrates that the $c$ axis of the magnetic easy axis is parallel to the specimen surface.

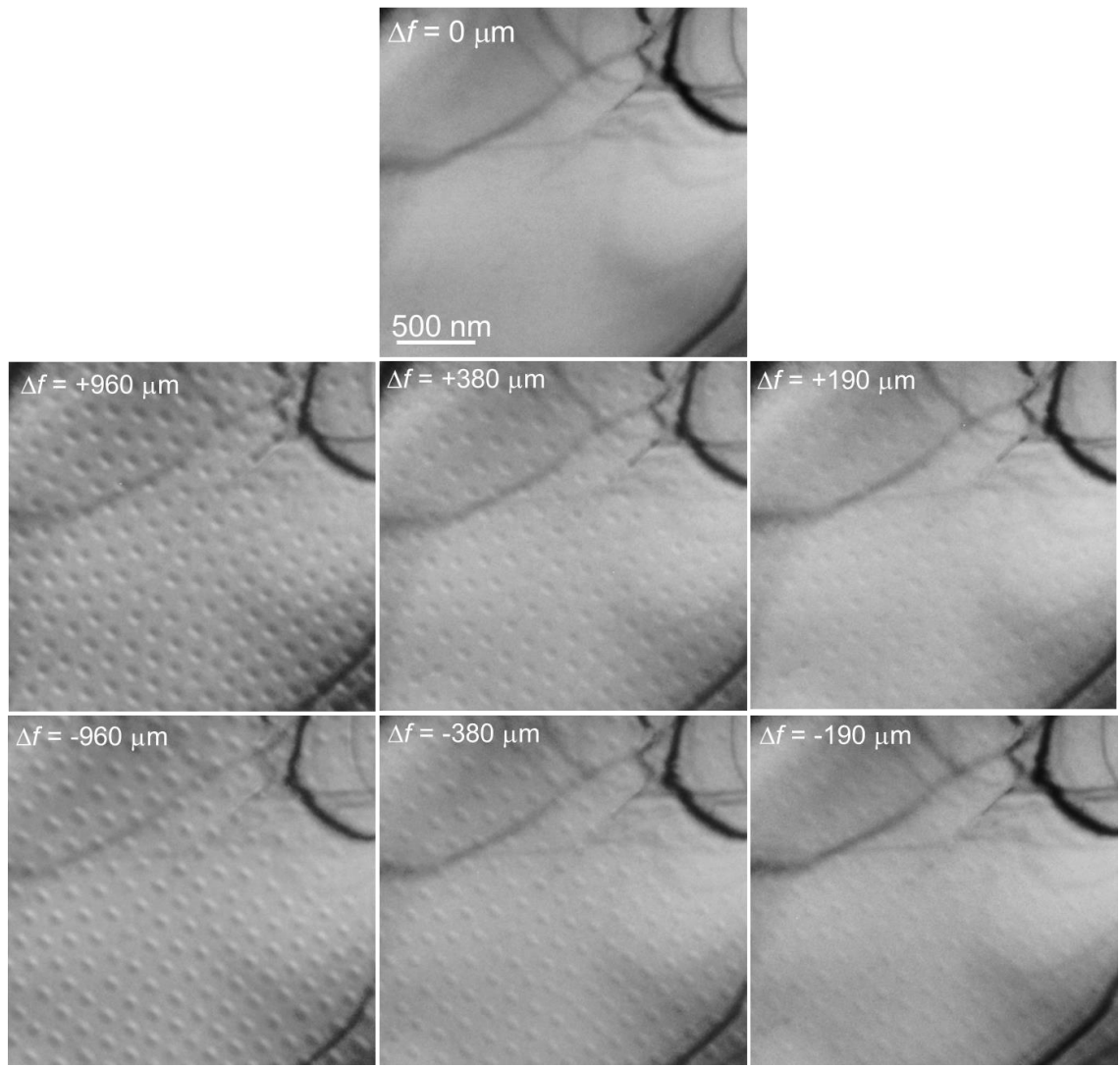

**Supplementary Figure 2. Through-focus images of magnetic bubbles comprising 180° domains (Biskyrmions).** The defocus value $\Delta f$ is shown in each Fresnel image. The area is the same as in Figure 3 of the main text. The applied magnetic field was 300 mT.

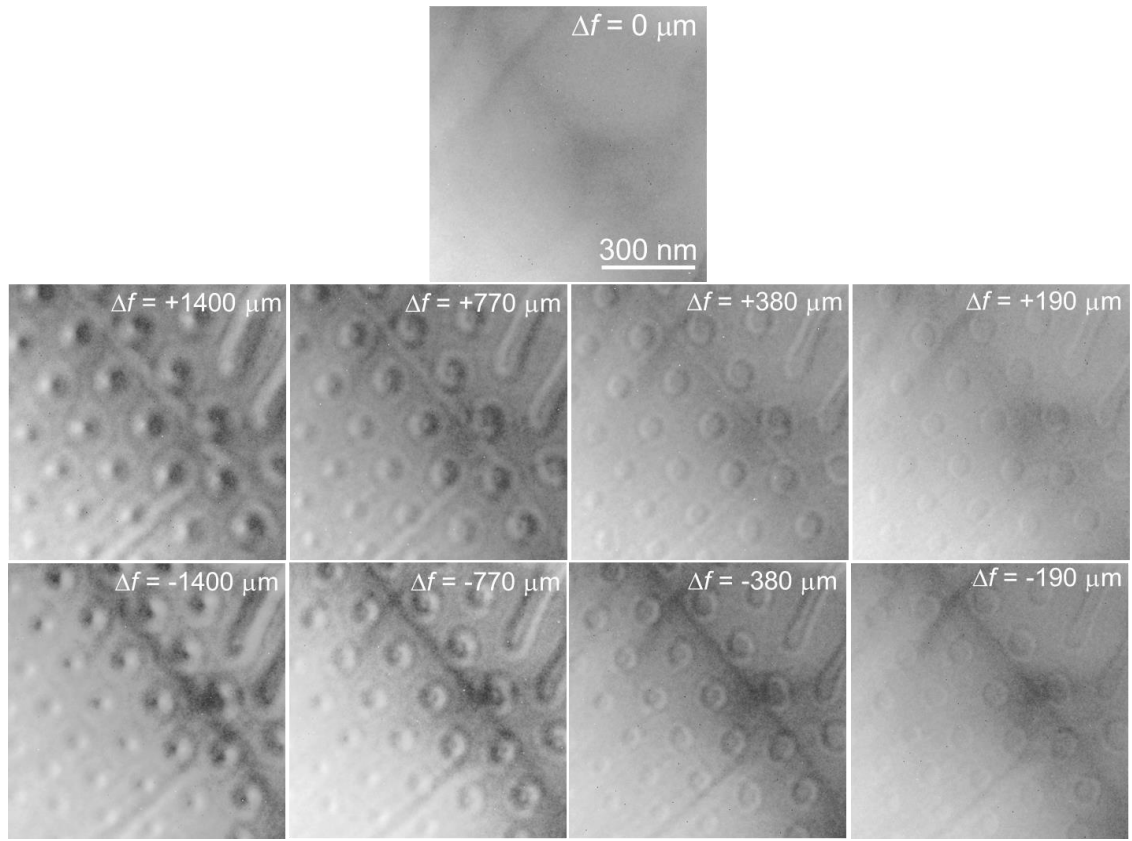

**Supplementary Figure 3. Through-focus images of type-II magnetic bubbles.** The defocus value Δ$f$ is shown in each Fresnel image. The area is the same as in Figure 4 of the main text. The applied magnetic field was 180 mT.